\documentclass{aa}
\usepackage{graphicx}
\usepackage{balance}
\newcommand{\teff}{T$_{\rm eff}$}

\begin{document}
\idline{1}{0}
\doi{0}
\def\year{2005}

\title{Li in NGC6752 and the formation of globular clusters
\thanks{Based on observations collected at the ESO VLT, Paranal Observatory,
Chile, Archive data of the Programme 165.L-0263}}

\subtitle{}
\author{L. Pasquini\inst{1}, P. Bonifacio\inst{2}, P. Molaro\inst{2}, P. Francois\inst{3}, 
F. Spite\inst{3}, R. G. Gratton\inst{4}, E. Carretta \inst{5}, B. Wolff\inst{1}}

\offprints{lpasquin@eso.org}

\institute{European Southern Observatory, Garching bei M\"unchen, Germany
\and
INAF--Osservatorio Astronomico di Trieste, Trieste, Italy
\and
DASGAL- Observatoire de Paris-Meudon, France
\and
INAF--Osservatorio Astronomico di Padova, Padova, Italy 
\and 
INAF--Osservatorio Astronomico di Bologna, Bologna, Italy }

\abstract{Li abundances for 9 Turnoff (TO) stars
of the intermediate metallicity cluster ([Fe/H]=--1.4) NGC6752 are presented.
The cluster is  known to show  abundance anomalies and anticorrelations 
observed  in both evolved and main sequence stars. We find that Li 
abundance  anticorrelates with Na (and N) and correlates with O 
in these Turn-Off  stars. For the first 
time we observe  Pop II hot dwarfs  
 systematically departing from the {\em Spite plateau}. 
The observed anticorrelations are in qualitative 
agreement with what is expected if the 
original gas were contaminated by Intermediate Mass  AGB - processed material. 
However, a quantitative comparison  shows   that none of the  
existing models can reproduce all the observations at once.
The very large amount of processed gas present in the cluster does
not imply a 'pollution', but rather that the whole 
protocluster cloud was enriched by a previous generation of  stars.
We finally note that the different  abundance patterns in NGC 6397 
and NGC 6752 imply different ejecta of the preenrichment composition 
for the two clusters. 
}
\authorrunning{L. Pasquini et al.}
\titlerunning{Li abundance in NGC~6752}
\maketitle


\section{Introduction}

With the advent of 8m telescopes we are  able to obtain 
high resolution and high quality spectra for stars belonging to the 
main sequence of globular clusters, which have 
allowed astronomers to derive accurate abundances for them.  
These abundances  have  set 
several limits  to the physics of stellar  
atmospheres, and have shed  some light on the long
debated problem of the origin  of  chemical anomalies 
in globular clusters (Thevenin et al. 2001, 
Gratton et al. 2001 (hereafter G01), James et al. 2004, Carretta et al. 2004). 
In this context Li abundance studies play  a special role, 
given the fragility of this element which can be easily destroyed in the
stellar interior. Indeed, while early Li studies in globular 
clusters  mostly concentrated  on the problems related to the primordial nature
of the Li {\em plateau} (Molaro and Pasquini, 1994, Pasquini and Molaro 1996,1997,
Delyiannis et al. 1995, Boesgaard et al. 1998), later studies have emphasized the role of Li 
for understanding the mixing phenomena in globular cluster stars or for
constraining the mechanism responsible for the chemical  pollution of the clusters 
(Castilho et al. 2000, Thevenin et al. 2001, Bonifacio et al. 2002, Grundahl et al. 2002). 
The only cluster studied in some detailed as faint as  the Turnoff 
 is the nearby, metal poor NGC6397, which exhibits  an impressively constant
Li abundance  among  TO stars, at the same level of the field stars 
{\em plateau} (Bonifacio et al. 2002). Since  NGC6397  shows   
chemical inhomogeneity  in the  oxygen abundance among main sequence stars 
and a  high nitrogen abundance, as recently reported by Pasquini et al. 
(2004), it is difficult to explain the  {\em plateau} Li abundance without a 
fine tuning.   The  proposed Li production 
from intermediate mass (IM) AGB stars should give  yelds very close to the 
values of primordial nucleosynthesis. The presence of a significant 
amount of  beryllium 
suggests that these IM-AGB stars  formed  very early after the big bang, 
and polluted the gas which was later exposed for several hundred million years 
to the galactic cosmic ray flux 
before the stars we now observed formed
(see the discussion in 
Ventura et al. 2001, 2002, Bonifacio et al. 2002, Pasquini et al. 2004). 
Since the case of NGC6397 would require a fine tuning between Li
production and destruction in the IM-AGB to reproduce exactly the Li {\em plateau} level observed, 
it is interesting to investigate the behaviour of Li in other clusters to 
test if they carry  the signature of AGB pollution. 
NGC6752 has  a metallicity of 
[Fe/H]=--1.43, and with a temperature  of about $\sim$6200 K, 
its TO stars  belong to the Li {\em plateau} (Spite and Spite 1982). 
NGC6752 is therefore an ideal cluster for this study. The cluster 
is one of the  prototypes of globular cluster with chemical anomalies,
where  the first   O-Na anticorrelation has been discovered among 
TO stars (G01).

\section{Sample selection and observations}

We selected the stars of the G01 sample. 
Stars numbers together with their photometric 
properties, atmospheric parameters, Li and Na 
abundances as derived in G01 are listed in Table 1. 
These stars are in the same 
metallicity and temperature scale as NGC6397 stars in G01. 
Although G01 showed that the temperature of the stars is 
compatible with a single value, since  Li abundance is 
very sensitive to  small differences in effective temperature, 
we also computed the effective temperature of each star by 
assuming the observed  photometric values, 
the reddening of E(b-y)=0.032 (Gratton et al. 2003 (G03)), 
a  metallicity of -1.43 as derived by G01  and the Alonso et al. 1996 scale. 
ESO archive was searched to identify UVES observations of 
similar stars, but no spectra of TO stars were available. 

Abundances were computed with the method outlined in Bonifacio et al. 
2002. Kurucz models were computed with the appropriate 
metallicity and temperature, and the Li abundances 
derived from the observed equivalent  widths (EWs). 
The typical error in the Li abundance is of 0.05 dex.
We do expect an error of up to 
0.1 in A(Li) \footnote{For chemical abundances we use the 
notation A(X)=log(X/H)+12.} 
when considering  possible uncertainties of up to 100 K in 
effective temperature. 
However, we shall consider that the  moderate reddening of the 
cluster (E(B-V)=0.040) and the use of the same temperature scale for all the stars 
implies that most of this uncertainty applies  to the  absolute 
value of  Li abundance, but much less 
to the results concerning the 
dispersion of  Li abundance in the cluster. 
The latter should be dominated by the photometric error 
in the b-y colour, which should not exceed 0.02. This  translates in an 
uncertainty of $\approx$ 60 K in the {\teff} or 0.05 dex in A(Li). 

\begin{table*}
\caption{NGC6752 stars, their atmospheric parameters and Li
abundances. Stars are ordered by increasing [Na/Fe] abundances.  
The first three columns report the observed photometric parameters.   
The G01 Temperature was of 6226  K and the surface gravity 
of log g = 4.28 for all stars.
The Li value in Column 6 and [Na/Fe] in column 7 are 
computed with these stellar parameters. Column 8 presents  {\teff} computed 
according  to the  Alonso scale, b-y colour and the reddening of E(b-y)= 0.032
(G03). The [Na/Fe] abundances of Column 9 and the Li abundances in 
Column  10 are computed adopting the last  temperatures.}

\begin{tabular}{lllllrlrlr}
 \hline
 Star   & V       & b-y     & c$_1$ &  Li E.W.  & A(Li)  & [Na/Fe]   &  {\teff} & [Na/Fe] & A(Li)   \\ 
        &         &         &       &  m\AA     & (G01)  & (G01)     &  K       & Alonso &  Alonso     \\ 
 \hline
 \hline
4428    & 17.142   & 0.366 & 0.307 &  42.8     & 2.50 $\pm$ 0.04    & --0.35     & 6013 & --0.28   &  2.35 \\
4383    & 17.112   & 0.360 & 0.284 &  35.9     & 2.41 $\pm$ 0.04    & --0.23     & 6043 & --0.17   &  2.28 \\
202316  & 17.275   & 0.354 & 0.237 &  31.7     & 2.35 $\pm$ 0.07    & --0.09     & 6064 & --0.04   &  2.23 \\
4341    & 17.149   & 0.345 & 0.324 &  29.4     & 2.31 $\pm$ 0.05    &  0.18     & 6168 &  0.20   &  2.26   \\
4458    & 17.155   & 0.343 & 0.335 &  21.2     & 2.14 $\pm$ 0.04    &  0.24     & 6188 &  0.25   &  2.11   \\
4661    & 17.216   & 0.342 & 0.281 &  20.9     & 2.14 $\pm$ 0.10    &  0.28     & 6170 &  0.30   &  2.10   \\
5048     & 17.284   &0.353 & 0.356 &  21.4     & 2.15 $\pm$ 0.09  &  0.37     & 6126 &  0.40   &  2.08     \\
4907     & 17.199   &0.354 & 0.306 &  17.9     & 2.06 $\pm$ 0.06  &  0.61     & 6096 &  0.65   &  1.97     \\
200613   & 17.198   & 0.377& 0.330  &  20.5     & 2.13 $\pm$ 0.06  &  0.64     & 5948 &  0.73   &  1.93    \\
Average  & 17.192   &      &   & 26.9      & 2.24             &           & 6091    &      &  2.15         \\
$\sigma$ & $\pm$0.06 &     &   & $\pm 8.5$ & $\pm$0.15        &           & $\pm$81 &      & $\pm$0.14     \\
\hline \hline
 \end{tabular}
\end{table*}

\begin{figure*}[ht]
\begin{center}
\resizebox{\hsize}{!}{\includegraphics{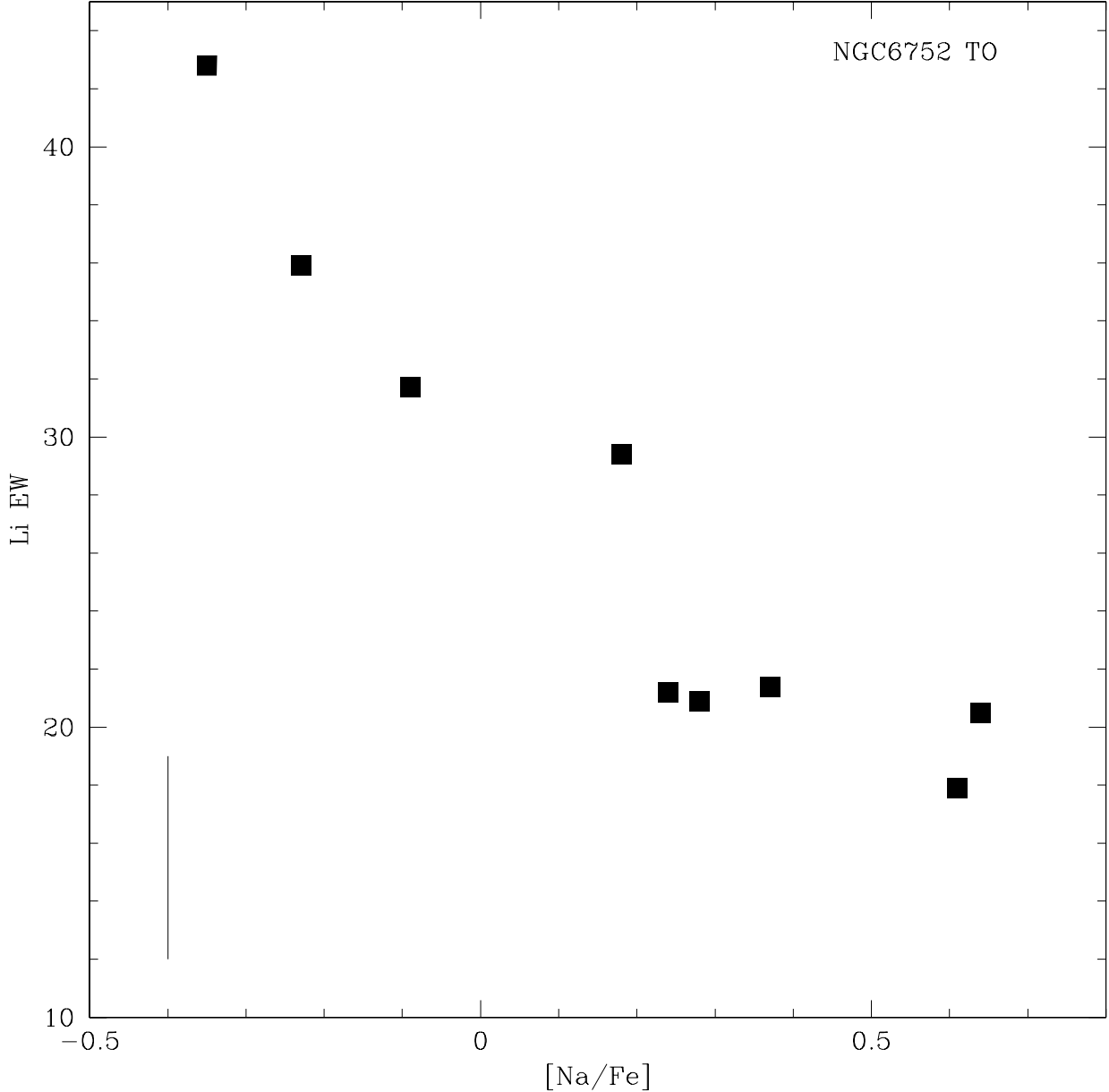}}
\end{center}
\caption{ Li Equivalent widths vs. [Na/Fe] for the NGC6752 stars.
The typical measurement error bar is given in the left bottom corner.  }
\end{figure*}

\begin{figure*}[ht]
\begin{center}
\resizebox{\hsize}{!}{\includegraphics{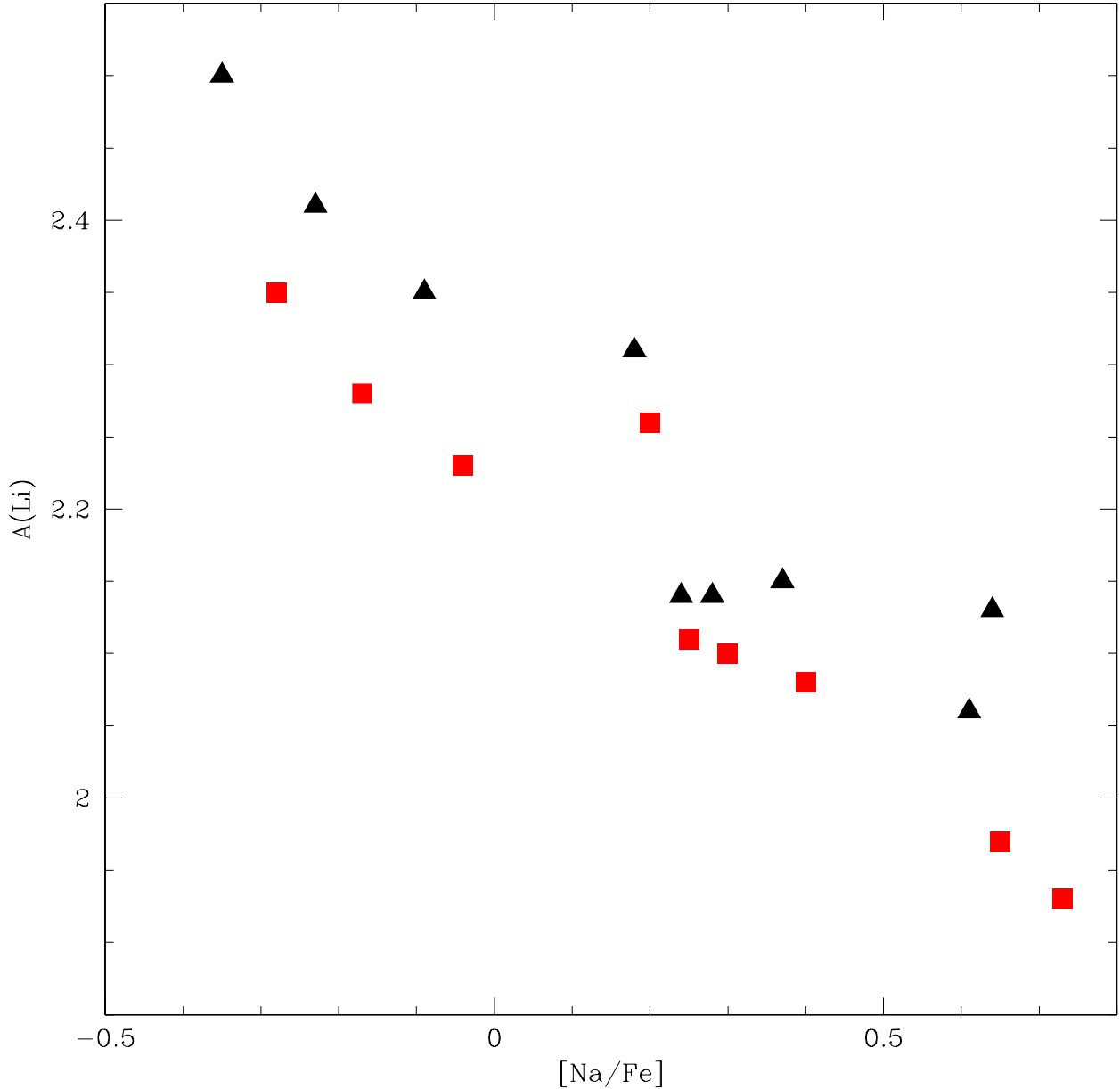}}
\end{center}
\caption{ Li abundances vs. [Na/Fe] for the sample stars. For the sake of completeness, 
abundance values for both temperature scales are plotted: filled triangles G01 scale;
filled (red) squares: Alonso scale.   } 
\end{figure*}

\begin{figure*}[ht]
\begin{center}
\resizebox{\hsize}{!}{\includegraphics[clip=true]{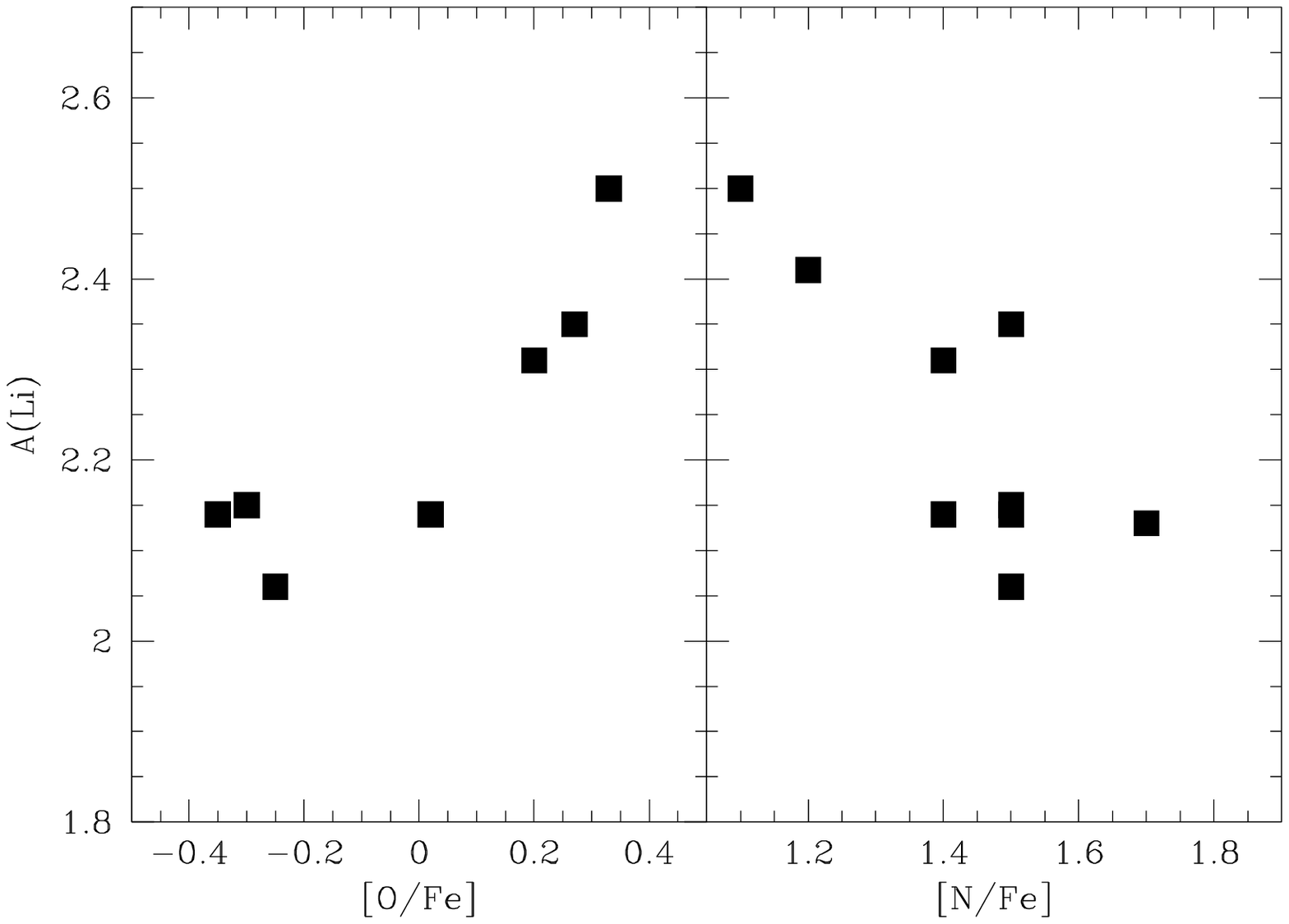}}
\end{center}
\caption{ Li abundances vs. [O/Fe] and vs. [N/Fe] 
for the sample stars. 
[O/Fe] and [N/Fe] values are from Table 1 of Carretta et al. 2005, 
who used the  G01 temperature scale.  } 
\end{figure*}

\section{Discussion}

The equivalent width measurements of Table 1 indicate  at first glance a 
variability of Li. 
We know, however, that to establish the real Li 
variations in a cluster  requires a proper analysis 
of the errors and of the additional, possible hidden biases introduced by the analysis 
method (see e.g. Bonifacio  2002). 
Since the stars are consistent  with a single effective temperature we  start the analysis
by considering the  equivalent widths only. The S/N ratio of the observations varies 
between 34 and 67 (per pixel) and the  errors estimated in the equivalent widths range between
1.8 m{\AA} for the best exposed spectra to 4 m{\AA} for those with lower S/N. 
(The relative S/N ratio  among the stars can be deduced by the errors in the Li abundances given in Column 6). 
The difference in Li equivalent width among different stars is between 5 to 10 times larger than 
the typical measurement error for any object.  
The average of the  equivalent widths is of 26.9 m\AA, 
with a $\sigma$ of 8.5 m\AA. The dispersion is  2-5 times larger than the 
measurement errors on the single spectra.
 In Figure 1  
 the Li equivalent widths vs. Na abundance are plotted. The figure  shows a clear anticorrelation between Li and Na, and the 
Kendall's $\tau$ test provides an anticorrelation probability of 99.78\%.

To exclude that Li variations could  be an artifact
produced by possible  temperature differences
among the stars, we 
recomputed  the Li abundances adopting the Alonso et al. {\teff} values  
given in column 8 of Table 1. The resulting Li abundances  
show a lower Li mean  level (this is not surprising, being the photometric scale 
135 K lower) and a Li scatter slightly  lower but comparable  with what 
obtained with the unique temperature hypothesis. 
This confirms the presence of the Li-Na anticorrelation. 

With  this  temperature scale  the [Na/Fe] 
abundances will also change. 
But, since [Na/Fe] increases by decreasing the temperature
( [Na/Fe]  increases
by 0.034 dex for a difference in temperature of -100 K, see G01)
while Li abundance is decreasing, 
the Li-Na relationship remains substantially unchanged, as can be seen 
in Figure 2 and by directly comparing the values tabulated in 
Table 1: the [Na/Fe] range spanned by the stars is 
about one order of magnitude, irrespective of the {\teff} scale used.

We therefore are confident we found evidence for the first time  
of Li - Na anticorrelation in a group of GC stars with characteristics 
(metallicity, temperature and gravity) 
close to those of the {\em Spite plateau}.

There is no general consensus about  the metallicity 
at which  the {\em plateau} ends in the field halo dwarfs. 
Bonifacio \& Molaro (1997) defined this edge to occur  around
[Fe/H]=--1.5, where the first signs of stellar Li depletion  
start to appear. NGC 6752 with [Fe/H]=-1.4 is near this edge, 
being in fact slightly more metal-rich. The metal enhancement is however 
so small (and also dependent on the zero point adopted) to make the 
belonging of these TO stars to the {\em plateau} unquestionable. 

We interpret the Li-Na anticorrelation  as evidence that the gas forming 
NGC 6752 has been contaminated by a  
previous population which is responsible for the chemical inhomogeneities.

We must analyze the extent to which our targets have been polluted by 
the processed material. This aspect, in turn, will also provide us with fresh 
information about the details of the cluster formation.

\subsection{Implications on cluster chemical anomalies and  cluster formation}

Figure 2  shows the behaviour of Li abundances vs. Na abundances in our stars. 
In NGC6752 Na anticorrelates with O (G01), and correlates with N
(Carretta et al. 2005); in Figure 3 the behaviour of Li with [O/Fe] 
and [N/Fe] is shown, confirming the 
correlation with [O/Fe], while the same anticorrelation observed for Na
is recorded for N. 
The most immediate explanation is that the  Na-poor
stars, which are also  O and Li rich,  
have  a composition close to the 'pristine' one, 
while the Na (and N) rich stars are progressively 
contaminated. 

An important point to recall is  that the CNO cycle, 
which makes the N overabundance and O underabundance, and 
the Ne-Na  cycle which produces the Na overabundance occur at very high temperatures, 
20-30 times higher than the $\sim$2.5 Million K  at which Li is destroyed. 
It is thus expected that in the places where these cycles occur, no Li 
is left.  If 'pristine' and 'processed' material are mixed, 
then Li, Na and N  are expected to show some anticorrelation, and 
Li and O some correlation.  

In figures 1 and 2 There are two additional relevant aspects to be considered. 
A first aspect concerns the stars with the higher Li. 
These stars show the lowest Na and highest O and most likely likely 
they are very 
little polluted by processed material.  If we take 
the Li abundance of these  stars on the G01 temperature scale at face values, 
their A(Li) is about 2.45, or  0.1 dex higher than the {\em plateau} level. 
We  note  that these values are also found in   NGC6397 stars
when adopting the same temperature scale (Bonifacio et al. 2002, 
their Column 7 in Table 2). We interpret therefore this higher Li 
abundance as entirely due to the use of the G01 temperature scale, which is
hotter than the  Alonso and  the Bonifacio et al. (2002)
temperature scales. 

A second aspect refers to the most Li-poor stars:
it is worth noticing that 
also in the most Li-poor stars  the Li 
line is always  detected, although at an abundance level of A(Li)$\sim$2, 
or 2-3 times lower than in
the stars with the highest Li content.  
The fact that some Li is preserved even in the most Na-rich stars 
confirms  that the observed chemical anomalies  
have not been  produced by the star itself, but rather 
that the gas was processed previously somewhere else. This was   
shown by G01, because TO stars should not reach 
temperatures so high to ignite the Na cycle. 
However, this behaviour is different from what observed in the
metal poor cluster NGC 6397, where A(Li)  is constant.

The  contamination can be obtained in different ways either through 
Bondi accretion or through a process involving  the whole protocluster cloud. 
We favour the latter  because if the chemical anomalies were 
limited to the external accreted layers of the star, they should be later washed out 
when the stars  undergo the first dredge up, (as  
happens for Li, cfr. Grundahl et al. 2002 ). This  is not the case, since 
these anomalies are observed all along the RGB (see e.g. Carretta et al. 2005).
The fact that Li is observed even in the most 'contaminated' stars 
implies, then, that some Li must have been created by the previous 
generation of (contaminant) stars. 

The second possibility however implies such a huge contamination 
of the protocluster cloud that probably 
'contamination' is not  the most appropriate term anymore. 

The anomalous abundances suggest a precise 
 composition of the contaminating gas. 
The maximum difference observed in the 
Na abundance is almost one order of magnitude, 
the one in the Oxygen abundance is of about  0.8 dex, in the 
Li abundance is  only of about a factor 2.5. 
At the same time the  other heavier elements remain unchanged, 
and in particular 
the accreted material was not enriched in s-process elements (James et al.  2004). 
This shows that the most polluted 
stars have accreted at least  90$\%$ of their  gas which was Na rich, 
with a Li content lower but close to primordial and a negligible content of  Oxygen.
If a large fraction of the stars'
mass is indeed made out by this processed material, it is likely that this is 
just the signature of a group of stars in a limited mass range. We can
 draw  a scenario where  the elements created by supernovae are well 
 mixed in the protocluster, while the 
products of stellar  winds, with  lower velocity, would be more inhomogeneous. 

The ejecta of the previous generation of  stars had an upper limit content of 
A(Li)$\sim$2.0; A(Na) of at least 
$\sim$5.4; A(O) of less than $\sim$7.0 and A(N)$\sim$7.9.  
The general behaviour is qualitatively consistent with the models by Ventura et al. 2001, 2002 
who predicted the Li-O correlation from an intermediate-mass AGB contamination 
 and that Li should not be destroyed completely. 

In more quantitative terms, there is a rather good agreement 
with the models published by 
Ventura et al. 2002 for  very low metallicity IM-AGB. Ventura et al predict,  
for a Z=0.0006 initial composition, 
Li abundance of the order of  A(Li)$\sim$1.5-2,  
Oxygen abundance of $\sim$6.5-7.4  and  N abundance of $\sim$6.9-8.3 . 
Nitrogen abundance in NGC6752 TO stars is enhanced (Carretta et al 2005) 
showing clear evidence for CNO processing.  
According to the models, the low Oxygen abundance 
provides a clear indication that  the generation of stars producing the 
chemical inhomogeneities in  NGC6752  could only originate from 
4-5 M$_{\odot}$ metal poor (with Z $<$ 0.0006) AGB  stars.

The relatively high Li, on the other hand, is predicted 
to be produced only 
by fairly low mass (3 solar masses) and 
relatively metal rich progenitors. Although a full, 
detailed modelling might change these results,  
our preliminary conclusions are that  the observed Oxygen and Li 
abundances seem incompatible with progenitors of one type. 
We note that other works found that the Oxygen - Na anticorrelation 
cannot be quantitatively explained by the present IM-AGB models 
(Denissenkov and Herwig 2003, 
Palacios et al. 2005, Ventura and D'Antona  2005). 
Uncertainties in fundamental aspects of AGB evolution such as 
mass loss rate and treatment of convection at present seem to hamper 
the generation of realistic predictions for low metallicity AGB stars, and 
we might be at the stage where observations such as those presented here 
will serve to constrain evolutionary models rather than the opposite. 

Another important aspect is to understand the difference between 
the  Li behaviour in NGC6397 and in NGC6752 if the IM-AGB scheme were acting in 
both clusters. A corollary implication would be that the  ejecta 
of the contaminants  of the two clusters had different chemical compositions. 
Following the same argument as above and taking the O and N data from 
Pasquini et al. 2004, for NGC6397 we expect 
ejecta which were more rich in Li (about 2 times, or A(Li)$\sim $2.3), 
with an  O-poor content of A(O) of less than $\sim$6.7,  
while they had about A(N)$\sim$7.3. 
As far  Na is concerned, the value measured by 
G01 in the TO stars is at the level of A(Na)$\sim$ 4.5, constant among all stars, 
but the analysis of the subgiants by Carretta et al. (2005) shows clear 
variations with values of A(Na) up to $\sim$4.8. 

We  finally comment that  in order to explain  the  chemical variations
observed in the AGB context, huge pollution is required. The two  stars 
n4907 and n200613 should have been formed by more than
 $\sim$90$\%$ of IM-AGB processed material. 
If our sample of stars is indicative of the cluster population, it would imply that 
a large fraction (say about a half)   of the gas which formed the stars we now
observe was indeed processed by the previous IM-AGB stars population. 
The actual cluster mass is about 2$\times$ 10$^5$  solar masses, 
therefore at least 10$^5$ solar masses were processed by IM-AGB stars, 
leading to a minimum of $\sim$3$\times 10^4$ IM-AGB stars 
to produce the observed anomalies.  
Since there is no hint of the presence of low-mass stars belonging to this
first generation, this  implies a flat-topped IMF. A parallel effect would be
He enhancement produced by this IM-AGB processing, which was analyzed by 
Ventura et al. 2002 and by D'Antona et al. 2003. 
In addition, a considerable 
number of remnants  should be present in the cluster. 
These white dwarfs might be, however, 
not easily detectable: they would be likely in the 
faint tail of the luminosity function
and, in addition, they might have been 
segregated during the  complex dynamical history of the cluster. 

\balance
Even if at present AGB stars remain  the most promising candidates,  
the problems encountered in explaining all the observed features lead some 
groups to look for alternative scenarios to explain  the observed abundance patterns: 
Yong et al. (2005) suggested the presence of a new process producing 
simultaneously light and s elements in globular clusters; 
Piotto et al. (2005) invoked the
possible presence of low mass SNae to explain the  He-rich main sequence of 
$\Omega$Cen, and SNe with extensive fall-back were invoked 
in various flavours to explain the abundance patterns of the most metal poor stars
(Umeda and Nomoto 2003,  Limongi et al. 2003). 
Massive stars able to eject light elements, while retaining
the heavy elements locked in the SN remnant are, in principle, 
attractive  candidates. However, when analysing possible scenarios, 
we encounter several problems: it is difficult, for instance, 
 to produce the very low oxygen observed, 
to locate a  process of Li production, and, given the 
enormous mass of processed material required, 
a very peculiar IMF must be postulated. 

Quantitative element analysis,  such as that presented here, provide the 
experimental framework for solving this interesting puzzle. 

%
 \begin{acknowledgements}
We  thank Gabriella Schiulaz for a careful reading of the manuscript. 
\end{acknowledgements}

\end{document}